\documentclass[conference]{IEEEtran}
\usepackage{cite}
\usepackage{amsmath,amssymb,amsfonts}
\usepackage{algorithmic}
\usepackage{graphicx}
\usepackage{hyperref}
\usepackage{textcomp}
\usepackage{xcolor}
\usepackage{float}
\usepackage{tabularx}
\usepackage{multirow, multicol}
\usepackage{todonotes}

\AtBeginEnvironment{acronym}{ 
    }
\usepackage[nolist]{acronym} 

\def\BibTeX{{\rm B\kern-.05em{\sc i\kern-.025em b}\kern-.08em
    T\kern-.1667em\lower.7ex\hbox{E}\kern-.125emX}}

\begin{document}

\title{Impact Assessment of Cyberattacks in Inverter-Based Microgrids} 

\author{
\IEEEauthorblockN{\textbf{Kerd Topallaj, Colin McKerrell, Suraj Ramanathan, Ioannis Zografopoulos}}
\IEEEauthorblockA{\textit{Engineering Department} \\
\textit{University of Massachusetts Boston}\\
Boston, MA, USA \\
\{kerd.topallaj001, colin.mckerrell001, suraj.ramanathan001, i.zografopoulos\}@umb.edu}
}

\maketitle

\begin{abstract}
In recent years, the evolution of modern power grids has been driven by the growing integration of remotely controlled grid assets. Although \ac{DERs} and \ac{IBR} enhance operational efficiency, they also introduce cybersecurity risks. The remote accessibility of such critical grid components creates entry points for attacks that adversaries could exploit, posing threats to the stability of the system.
To evaluate the resilience of energy systems under such threats, this study employs real-time simulation and a modified version of the IEEE 39-bus system that incorporates a \ac{MG} with solar-based \ac{IBR}. 
The study assesses the impact of remote attacks impacting the \ac{MG} stability under different levels of \ac{IBR} penetrations through \ac{HIL} simulations. Namely, we analyze voltage, current, and frequency profiles before, during, and after cyberattack-induced disruptions. The results demonstrate that real-time \ac{HIL} testing is a practical approach to uncover potential risks and develop robust mitigation strategies for resilient \ac{MG} operations.
\end{abstract}

\vspace{0.4cm}

\begin{acronym}[IES-DSE] 
\acro{AI}{Artificial Intelligence}
\acro{AGC}{Automatic Generation Control}
\acro{AMI}{Advanced Metering Infrastructure}
\acro{API}{Application Programming Interface}
\acro{ADMS}{Advanced Distribution Management Systems}
\acro{CNN}{Convolutional Neural Network}
\acro{CNT}{Complex Network Theory}
\acro{CPPS}{Cyber-Physical Power System}
\acro{CIM}{Common Information Model}
\acro{CGMES}{Common Grid Model Exchange Standard}
\acro{CPS}{Cyber-Physical System}
\acro{CHIL}{Controller Hardware-in-the-Loop}
\acro{DERs}{Distributed Energy Resources}
\acro{DoS}{Denial-of-Service}
\acro{DL}{Deep Learning}
\acro{DOE}{Department of Energy}
\acro{DT}{Digital Twins}
\acro{DSE}{Dynamic State Estimation}
\acro{DSO}{Distribution System Operator}
\acro{EMS}{Energy Management System}
\acro{ENTSO-E}{European Network of Transmission System Operators for Electricity}
\acro{EV}{Electric Vehicles}
\acro{EMT}{Electromagnetic Transient}
\acro{FACTS}{Flexible Alternating Current Transmission Systems}
\acro{FDIA}{False Data Injection Attack}
\acro{FNCS}{Framework for Network Co-simulation}
\acro{GNN}{Graph Neural Network}
\acro{GOOSE}{Generic Object Oriented Substation Event}
\acro{GFM}{Grid-Forming}
\acro{GFL}{Grid-Following}
\acro{HIL}{Hardware-in-the-Loop}
\acro{HELICS}{Hierarchical Engine for Large-scale Infrastructure Co-simulation}
\acro{HV}{High-Voltage}
\acro{HVDC}{High-Voltage Direct Current}
\acro{HPC}{High Performance Computing}
\acro{IBR}{Inverter-Based Resources}
\acro{ICT}{Information and Communication Technology}
\acro{IED}{Intelligent Electronic Device}
\acro{IoT}{Internet of Things}
\acro{IP}{Internet Protocol}
\acro{IT}{Information Technology}
\acro{ISO}{Independent System Operator}
\acro{KNN}{K-nearest Neighbor}
\acro{LFC}{Load Frequency Control}
\acro{LMP}{Locational Marginal Price}
\acro{ML}{Machine Learning}
\acro{MG}{Microgrid}
\acro{MLP}{Multilayer Perceptron}
\acro{MITM}{Man-in-the-Middle}
\acro{MDP}{Markov Decision Process}
\acro{MPS}{Mobile Power Resources}
\acro{MTD}{Moving Target Defenses}
\acro{MV/LV}{Medium-to-low Voltage}
\acro{NASPI}{North American Synchrophasor Initiative}
\acro{OSI}{Open Systems Interconnection}
\acro{OT}{Operational Technology}
\acro{PCA}{Principal Component Analysis}
\acro{PCC}{Point of Common Coupling}
\acro{PEPS}{Power Electronic-dominated Systems}
\acro{PLC}{Power Line Communication}
\acro{PMU}{Phasor Measurement Units}
\acro{PNNL}{Pacific Northwest National Laboratory}
\acro{PHIL}{Power Hardware-in-the-loop}
\acro{PSS}{Power System Stabilizer}
\acro{QoS}{Quality of Service}
\acro{RL}{Reinforcement Learning}
\acro{SCADA}{Supervisory Control and Data Acquisition}
\acro{SE}{State Estimation}
\acro{SGAM}{Smart Grid Architecture Model}
\acro{SDN}{Software-defined Network}
\acro{SSE}{Static State Estimation}
\acro{STATCOM}{Static Synchronous Compensator}
\acro{SV}{Sample Values}
\acro{SVM}{Support Vector Machine}
\acro{TF}{Task Force}
\acro{TnD}{Transmission and Distribution}
\acro{TSO}{Transmission System Operator}
\acro{t-SNE}{t-distributed Stochastic Neighbor Embedding}
\acro{VPP}{Virtual Power Plant}
\acro{WAN}{Wide-area Network}
\acro{WAMS}{Wide-area Measurement Systems}
\acro{WADC}{Wide-area Damping Control}
\acro{WAMPAC}{Wide-area Monitoring, Protection, and Control}
\acro{PV}{photovoltaic}
\acro{PCB}{Printed Circuit Board}
\acro{SCADE}{Supervisory Control and Data Acquisition}
\acro{UDP}{User Datagram Protocol}
\acro{GPIO}{General-Purpose Input/Output}
\acro{CB}{Circuit Breaker}
\acro{DER}{Distributed Energy Resource}
\acro{SO}{ System Operators}
\acro{NREL}{National Renewable Energy Lab}
\acro{BESS}{Battery Energy Storage System}
\acro{UV}{Undervoltage}
\acro{MPPT}{Maximum Power Point Tracking}
\end{acronym}

\acresetall

\vspace{-3mm}
\section{Introduction} \label{s:intro}
The growing penetration of \ac{DERs} -- such as \ac{PV} arrays, wind turbines, and energy storage systems -- requires new approaches to maintain grid reliability and stability. The \ac{MG} concept has emerged as a key solution for integrating and managing both renewable and non-renewable \ac{DERs}~\cite{b1}. According to the \ac{NREL} definition, a \ac{MG} is “\textit{a group of interconnected loads and \ac{DERs} that acts as a single controllable entity with respect to the grid}”~\cite{b8}. Furthermore, \ac{MG}s can operate in both grid-connected and islanded modes, exchanging power with the main grid or operating autonomously to support local loads. \looseness=-1

This flexibility makes \ac{MG}s essential components for maintaining power system stability during grid disturbances resulting from accidental events, e.g., faults, or malicious incidents, e.g., cyberattacks. A \ac{MG}'s ability to coordinate generation and demand at the local level enhances resiliency, reduces operational costs, and can defer transmission and distribution network expansion plans. Furthermore, \ac{MG}s offer \ac{SO} the flexibility to respond to rapid fluctuations in on-site demand and supply by supporting high shares of \ac{IBR} and enabling decentralized control. \looseness=-1

As the integration of \ac{DERs} and \ac{MG}s continues to grow, ensuring their secure and resilient operation under both normal and disruptive conditions becomes increasingly critical. One of the primary areas of interest involves assessing the performance of \ac{MG} under adverse conditions, such as cyberattacks or unintentional faults, and examining their potential to trigger forced islanding events \cite{zografopoulos2025event}. These islanding transitions sectionalize \ac{MG}s from the main grid at their \ac{PCC}, which is typically controlled by a \ac{CB}. 
Rapid shifts between grid-connected and islanded modes can induce transient instability, frequency deviations, and voltage fluctuations, potentially compromising system reliability \cite{zografopoulos2021security}. Traditional testing methods focusing on offline simulations are often unable to capture real-time dynamic phenomena, while experimenting on the actual power systems or even smaller deployments is cost-prohibitive and could raise safety issues. As a result, it is essential to experiment with high-fidelity system models that respect the mission-critical and time-sensitive nature of the power system's critical infrastructure, without impacting actual grid operations\cite{katuri2023experimental}. \looseness=-1

Recent advances in real-time \ac{HIL} simulation offer a powerful solution to assess \ac{MG} behavior before field deployment. By integrating power system models with external hardware, \ac{HIL} enables realistic testing of operational stability and cyber-threats in a controlled environment, i.e., the cyber-physical testbed. Unlike purely software-based simulations, \ac{HIL} provides real-time feedback by allowing interaction with physical controllers, inverters, and protection devices. Additionally, \ac{HIL} methods reduce operational risks and enhance grid security by detecting vulnerabilities before real-world implementation.\looseness=-1

The contributions of this work are the following:
\begin{itemize}
    \item We combine essential power system assets and a \ac{MG} into an integrated \ac{TnD} model. For the transmission-level system we use the IEEE 39-bus system, while the \ac{MG} is comprised of a \ac{PV} farm complemented by synchronous generation. 
    \item We study the impact of sophisticated cyberattacks that, after identifying an anomalous grid condition, e.g., fault, they rapidly toggle the \ac{CB} at the \ac{PCC}, switching the \ac{MG} between islanded and grid-connected modes, and stressing the \ac{MG}'s capacity to maintain stability.\looseness=-1
    \item We present real-time simulation results 
    to illustrate the impact of different \ac{IBR} penetration levels on nominal operations and their potential to exacerbate grid instability.\looseness=-1
\end{itemize} 

\begin{figure}[t]
    \centering    \includegraphics[width=0.35\textwidth]{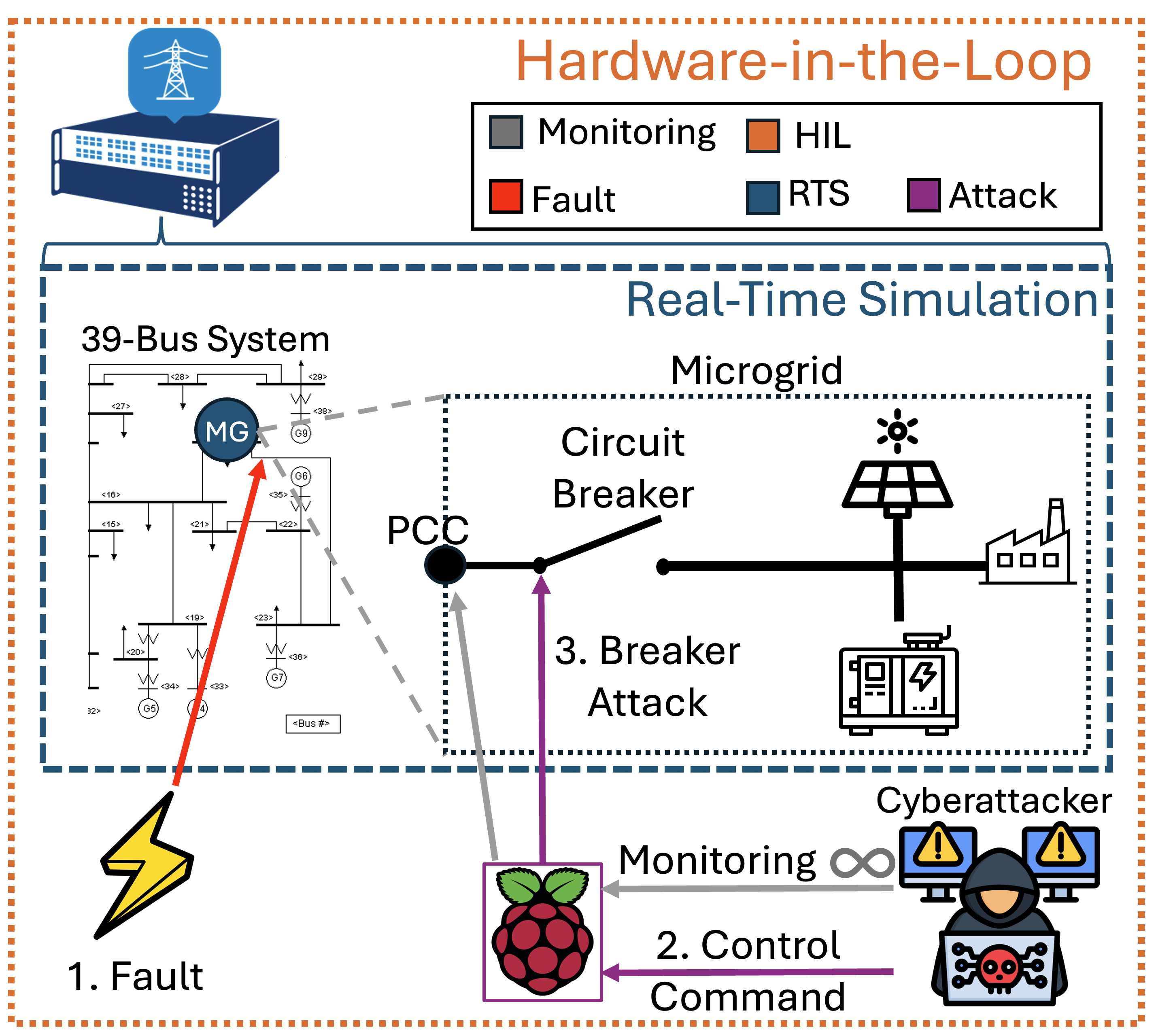} 
    \caption{System overview and attack methodology.}
    \label{fig:overview}
    \vspace{-3mm}
\end{figure}

An overview of the developed \ac{TnD} system and the cyberattack kill chain is shown in Fig.~\ref{fig:overview}. The remainder of this paper is organized as follows. Section~\ref{s:method} outlines the methodology, including the \ac{MG} system model and the implementation of cyberattack scenarios. Section~\ref{s:results} presents the simulation configuration and results under various attack conditions and generation mixes, analyzing their impact on system stability, frequency response, and voltage waveforms. Finally, Section~\ref{s:conclusion} concludes the paper and discusses directions for future work. \looseness=-1

\section{Methodology} \label{s:method}
The following subsections detail the power system model and the modifications performed to the IEEE 39-bus system to integrate the inverter-based \ac{MG}. We also outline the adversary model and cyberattack assumptions adopted in this study. \looseness=-1 

\subsection{System Model}

\begin{figure}[t]
    \centering    \includegraphics[width=0.35\textwidth]{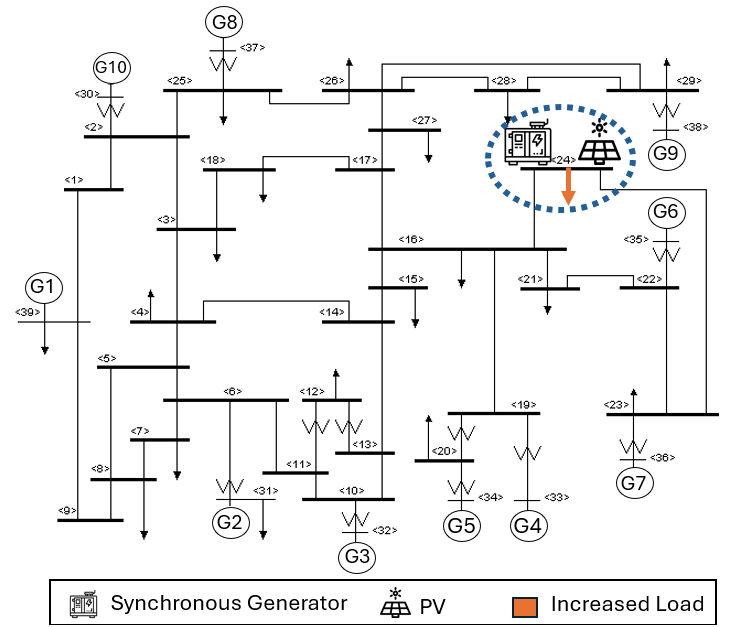} 
    \caption{Modified IEEE 39-bus transmission system.}
    \label{fig:39bus}
    \vspace{-3mm}
\end{figure}

This study uses the IEEE 39-bus transmission system to evaluate the performance and stability of an autonomous \ac{MG}. 
As shown in Fig.~\ref{fig:39bus}, the system comprises 10 synchronous generators, 34 transmission lines, 12 transformers, and 19 aggregated loads~\cite{b9}. The original New England system includes only synchronous generators. However, inverter-based \ac{DERs} (specifically a \ac{PV} farm), which lack natural inertia are integrated alongside conventional sources to reflect evolving generation portfolios. Different \ac{PV} penetration levels are examined in Section \ref{s:results} to assess their impacts on system stability.\looseness=-1

To reflect the \ac{MG}'s behavior in real-time during islanded operation, the \ac{PV} inverter is configured to operate in a \ac{GFM} fashion. \ac{GFM} inverters can independently regulate voltage and generate their own frequency reference, making them suitable for autonomous operation. 
Furthermore, bus 24 is selected as the \ac{MG} interconnection point, as shown in Fig.~\ref{fig:39bus} (blue circle). The additions of \ac{PV} penetration and a synchronous generator are also connected to bus 24. Although bus 24 is not directly connected to any generator, it is electrically adjacent to bus 23 and G7. This location could become a prominent target for an attacker aiming to propagate impacts to adjacent generators and loads. \looseness=-1

To evaluate the impact of adverse events on a stressed operational scenario, the load demand at bus 24 is increased by 20\%, indicated by the orange arrow at the bus in Fig.~\ref{fig:39bus}. This could represent residential or industrial load excursions during abnormal conditions, enabling the evaluation of weakly-connected \ac{MG} performance under high-loading conditions. 
Additionally, a single-phase-to-ground fault is introduced at bus 24 to assess the \ac{MG}'s ability to sustain the local loads during fault conditions. Without localized generation, such faults can lead to instability. However, with the \ac{MG} in place, the grid's post-fault dynamic behavior should be analyzed. \looseness=-1 

\subsection{Adversary Model}
The adversary model defines the attacker's capabilities, knowledge, and access~\cite{b11}. In this work, the adversary is assumed to have partial system knowledge, specifically, awareness of the remotely accessible communication interface of the \ac{CB}. This reflects a gray-box threat model in which the attacker lacks complete visibility into the bulk power system but understands how to launch targeted attacks against the \ac{MG}.

The adversary can remotely access and control a Raspberry Pi used as the cyber interface to the \ac{CB} that connects the \ac{MG} with the rest of the system, as seen in Fig. \ref{fig:overview}. This device enables remote attacks, such as timed \ac{CB} switching attacks, delivered during vulnerable conditions, e.g., grid faults or peak loading. The attacker aims to destabilize the \ac{MG} by forcing repeated transitions between grid-connected and islanded modes, potentially causing cascading failures and blackouts. By timely coordinating their attack, the risk for grid destabilization could increase during abnormal events. Thus, the adversary aims to leverage such a condition, i.e., during a single-phase-to-ground fault at bus 24, to maximize their impact on the system.\looseness=-1 

The attacker could be classified as a Class I adversary, as defined in~\cite{b11}. While they possess moderate resources and remote access to the \ac{CB}, their ability to carry out the attack covertly is limited. Although a single unauthorized transition may not trigger system-wide consequences, repeated and intentional switching at the \ac{MG}'s \ac{PCC} would likely be flagged by system operators as suspicious activity.

\subsection{Attack Methodology}
The attack model describes how a system vulnerability could be exploited to become a system-level threat~\cite{b13}. In our case, the attacker targets the \ac{CB} at the \ac{MG} \ac{PCC}, aiming to trigger unintentional islanding conditions. The primary vulnerabilities of the \ac{MG} lie in the insecure communication interfaces used by \ac{SO} to issue grid islanding commands by tripping the \ac{CB} at the \ac{PCC} \cite{b13, zografopoulos2021security}. Furthermore, sophisticated attackers can leverage the increased reliance of \ac{MG}s on predominantly \ac{IBR}-based generation to maximize their attack impacts since, unlike synchronous generators, \ac{IBR} resources lack inertia, making them unable to ``absorb" sudden disturbances.  
Thus, the \ac{CB} becomes a high-impact target given its remote accessibility and limited built-in security. Once it is compromised and maliciously operated, it can jeopardize system stability, as we demonstrate in Section \ref{s:results} \cite{kertzner2022cja}.

To exploit the identified vulnerabilities, the attack is carried out using a Raspberry Pi, which transmits unauthorized actuation signals to the  \ac{CB}, modeled in the OPAL-RT real-time simulation environment. The attack orchestration, shown in Fig. \ref{fig:overview}, includes the following stages. Under normal system conditions, the attacker employs the Raspberry Pi to passively monitor critical grid parameters, such as voltage and frequency, without initiating any active interference. This phase aims to collect system measurements and establish a baseline understanding of the grid's behavior. The attack is initiated once the adversary identifies an abnormal operating condition, such as a fault. The detection of abnormal grid conditions, achieved by closely analyzing the grid's real-time measurements, serves as the trigger for the attack.

Once an abnormal scenario is detected, the attacker orchestrates the attack. This involves overriding the legitimate control logic of the system by issuing malicious commands to the \ac{CB} (either to open or close it), thereby disrupting the grid's functionality \cite{mitreT0855}. The attacker can manipulate the \ac{CB} to isolate a portion of the grid through a single actuation, commonly referred to as a forced islanding attack (\emph{Scenario 1} in Table~\ref{tab:scenarioInfo}). Alternatively, the attacker may repeatedly issue commands to connect/disconnect the \ac{CB} multiple times, creating a switching attack (\emph{Scenario 2}). Following these steps, the attacker aims to disrupt the power grid, causing potential operational and reliability consequences.\looseness=-1
\section{Simulation Results} \label{s:results}

The following subsections delineate the experimental setup and outline the various simulation scenarios analyzed to evaluate the impact of cyberattack-induced islanding. 

\subsection{Experimental Setup} 

For our experiments, we utilize the IEEE-39 bus model, which has been modified to incorporate a \ac{MG} (Fig. \ref{fig:39bus}).  
The integrated \ac{TnD} model is developed using MATLAB Simulink and Simscape Electrical on a Windows-based workstation. This \ac{TnD} model is deployed onto the real-time simulator (OPAL-RT OP4610XG) to enable time-synchronized execution and interaction between the real-time environment and the external control node (Raspberry Pi 4) which has been maliciously compromised. \looseness = -1

An overview of the experimental setup is illustrated in Fig.~\ref{fig:overview}, where the Raspberry Pi is configured as an external, remotely accessible cyberattack vector. Communication between the real-time simulator and the Raspberry Pi is established using \ac{UDP}. This network configuration enables the transmission of real-time data and control signals between the two devices. 

\begin{table}[t!]
\small
    \setlength{\tabcolsep}{1.2pt}
    \centering
    \caption{Cyberattack Test Cases Information}
    \label{tab:scenarioInfo}
    
    \renewcommand{\tabularxcolumn}[1]{m{#1}}

    \begin{tabularx}{\linewidth} { 
      || >{\hsize=1.2\hsize\textwidth=\hsize\raggedright\arraybackslash}X 
      | >{\hsize=1.0\hsize\textwidth=\hsize\centering\arraybackslash}X
      | >{\hsize=0.8\hsize\textwidth=\hsize\centering\arraybackslash}X || }
      
     \hline \hline
     {MG Generation} & {{\textbf{Scenario 1}}} & {\textbf{Scenario 2}} \\ 
     \hline
     {\textbf{System I:~} 150MW PV \& 150MW Synchronous} &  \multirow{2}{8em}{Islanding at $t=1$s, reconnection at $t=1.5$s}  & \multirow{2}{7em}{\ac{CB} switching (6 times) between $t=1$s and $1.5$s}  \\\cline{1-1}
     
    {\textbf{System II:~} 210MW PV \& 90MW Synchronous} &  &   \\\cline{1-1}

    \hline \hline
    \end{tabularx}
\vspace{-3mm}
\end{table}

\subsection{Cyberattack Test Cases}
In Table \ref{tab:scenarioInfo} we summarize the specifics of the four different simulation test cases used in this work. In \emph{System I}, the 300\,MW of power generated in the \ac{MG} is evenly distributed (50\% -- 50\% split) between \ac{PV} and synchronous generation, while in \emph{System II}, the \ac{MG} operates with 70\% \ac{PV} generation (210\,MW) and 30\% synchronous generation (90\,MW). Each of the aforementioned test systems is then examined under two distinct scenarios. In \textit{Scenario 1} (\emph{single forced islanding}), the attacker issues two commands: the first trips the \ac{CB} at the \ac{PCC}, isolating the \ac{MG} from the main grid, and the second re-closes the \ac{CB}, restoring grid connection. In \textit{Scenario 2} (\emph{\ac{CB} switching attack}), the attacker rapidly toggles the \ac{CB}, causing the \ac{MG} to oscillate between islanded and grid-connected modes for three consecutive times.  

\subsubsection{System I -- Balanced \ac{IBR} and Synchronous \ac{MG} Generation}
\begin{figure}[t]
    \centering    \includegraphics[width=0.45\textwidth]{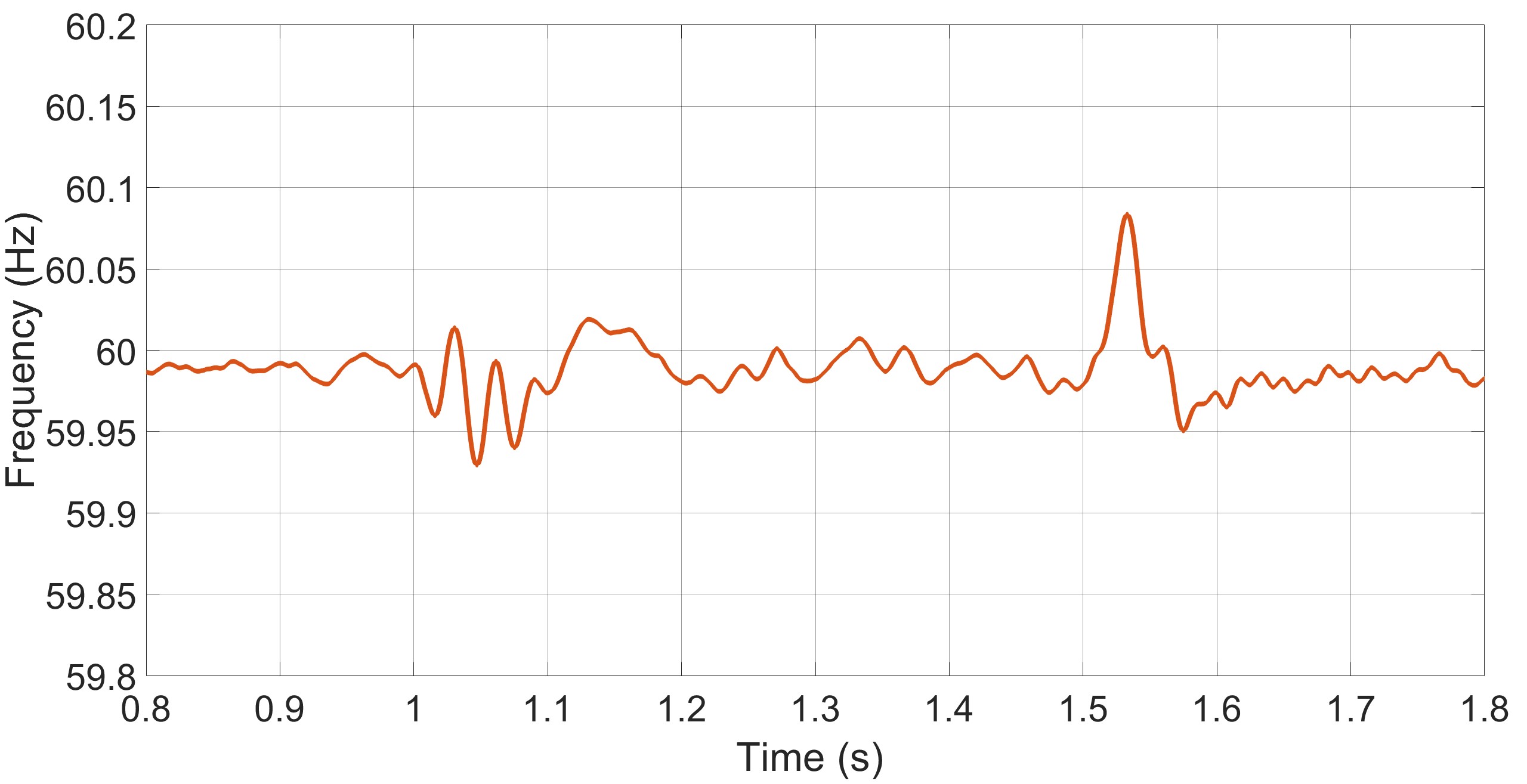}      \vspace{-3mm}
    \caption{Frequency response at the \ac{MG} (Bus 24) during forced islanding at $t=1$\,s and reconnection at $t=1.5$\,s.}
    \label{fig:freq_response_1}
\end{figure}

The following scenarios evaluate the stability of the \ac{MG} with balanced \ac{PV} and synchronous generation.
\paragraph{Single Forced Islanding Scenario}  
The attacker trips the \ac{CB} once to island the \ac{MG}. Fig.~\ref{fig:freq_response_1} shows the frequency response at the \ac{MG} at bus 24. According to the IEEE 1547 standard, the frequency should remain between the over-frequency threshold (OF1) of 61\,Hz and the under-frequency threshold (UF1) of 58.5\,Hz \cite{b10}. \looseness=-1

At $t\approx1$\,s, when the \ac{CB} opens, the frequency dips slightly and exhibits small oscillations below 60\,Hz. These oscillations remain limited, indicating that local generation stabilizes quickly in islanded mode. At $t \approx 1.5$\,s, reconnection causes a transient spike (60.08\,Hz) followed by a dip below 59.96\,Hz before settling near the nominal frequency. This larger deviation reflects the challenge of synchronizing two previously decoupled systems, i.e., the main grid and \ac{MG}. Overall, reconnection induces greater frequency swings than disconnection, but the \ac{MG} stabilizes within one second. Although the islanded frequency remains within 0.04\,Hz of the nominal range, some frequency excursions still exist as a byproduct of the reduced inertia within the \ac{MG} as the synchronous generator only supports half of the \ac{MG}'s load demand. \looseness=-1

\begin{figure}[t]
    \centering
    \includegraphics[width=0.45\textwidth]{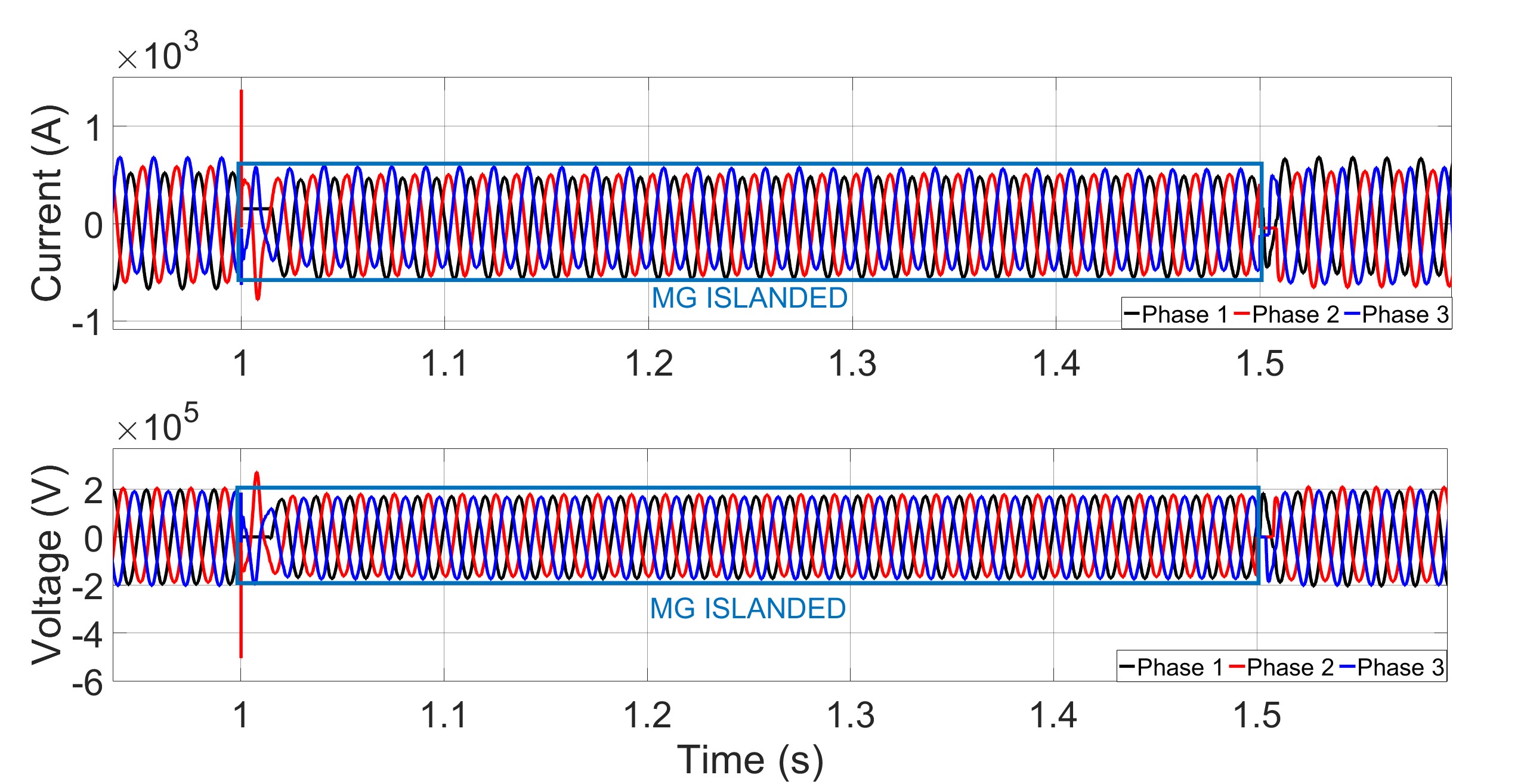}      \vspace{-3mm}
    \caption{Voltage and current waveforms at the \ac{MG} (Bus 24) during forced islanding at $t=1$\,s and reconnection at $t=1.5$\,s.} 
    \label{fig:islanding_waveforms_1} \vspace{-2mm}
\end{figure}

Fig.~\ref{fig:islanding_waveforms_1} shows the voltage and current waveforms at the \ac{MG} at bus 24. Upon islanding, both quantities decrease slightly in amplitude, but remain sinusoidal and stable, indicating successful local power delivery. Upon reconnection, they return to nominal levels after a brief transient spike, demonstrating fast synchronization with minimal disruptions.

\begin{figure}[t]
    \centering
    \includegraphics[width=0.45\textwidth]{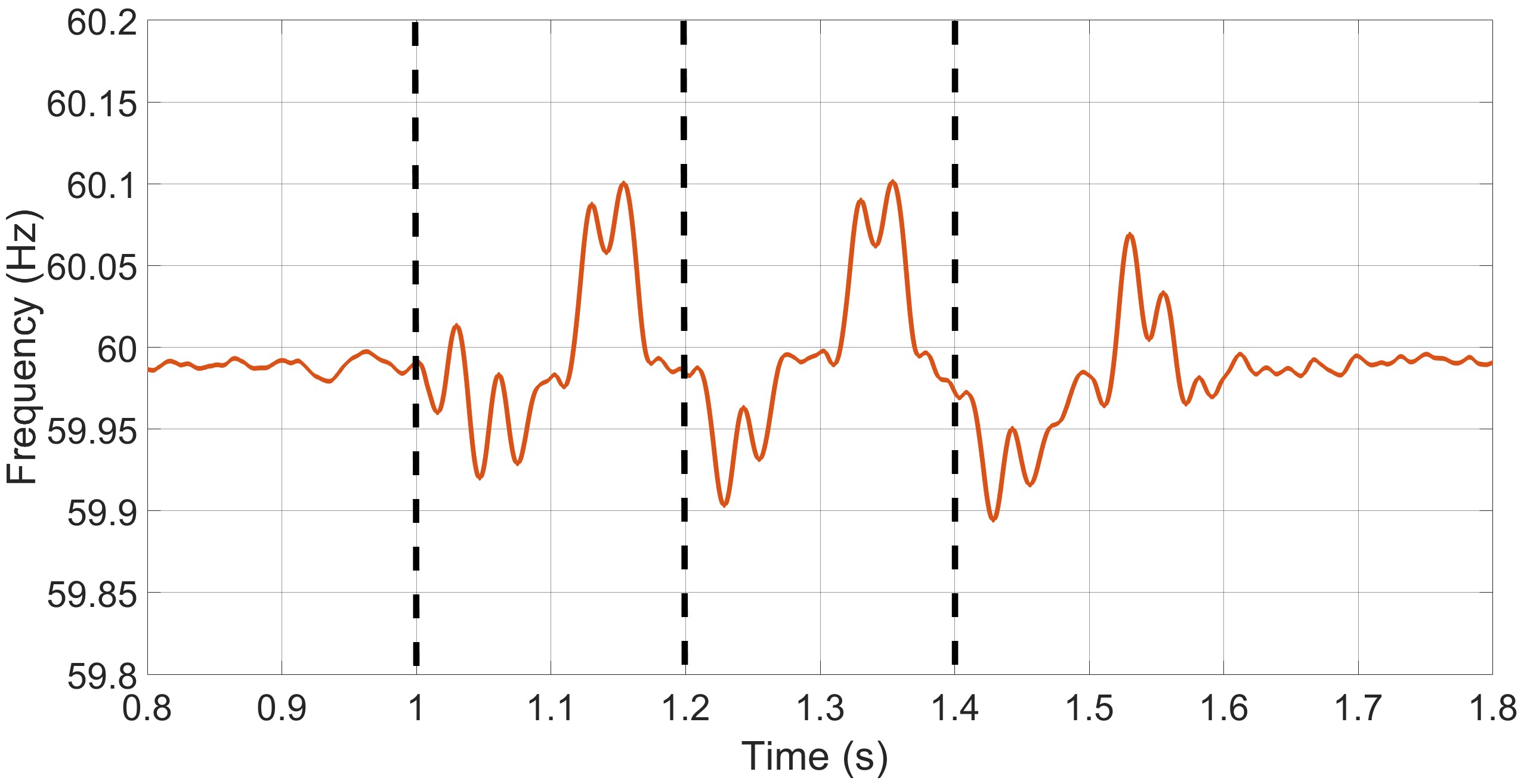}       \vspace{-3mm}
    \caption{Frequency response at the \ac{MG} (Bus 24) during switching attacks happening at $0.1$\,s intervals from $t=1$\,s to $t=1.5$\,s.}
    \label{fig:frequency_response_2}
\end{figure}

\paragraph{\ac{CB} Switching Attack Scenario}  
In this case, the attacker switches the \ac{CB} on and off three times. Fig.~\ref{fig:frequency_response_2} presents the frequency response at the \ac{MG}. The \ac{CB} is opened every 0.2 seconds between $t=1.0$\,s and $t=1.5$\,s and is closed 0.1 seconds after each opening. In Fig.~\ref{fig:frequency_response_2}, the dashed black lines indicate \ac{CB} openings. Each \ac{CB} opening results in a sharp frequency drop, and since the \ac{CB} closes before full recovery, the effects compound with each attack cycle. Subsequent islanding events deepen the frequency dips, threatening \ac{MG} stability, and upon each reconnection, the frequency briefly spikes to around 60.1\,Hz. Lastly, during the final reconnection at $t=1.5$\,s, the frequency spike is approximately 0.04\,Hz lower than earlier events, suggesting that the synchronous generation of the main system provides most of the inertia required to attenuate repeated disturbances. \looseness=-1

\begin{figure}[t]
    \centering
    \includegraphics[width=0.45\textwidth]{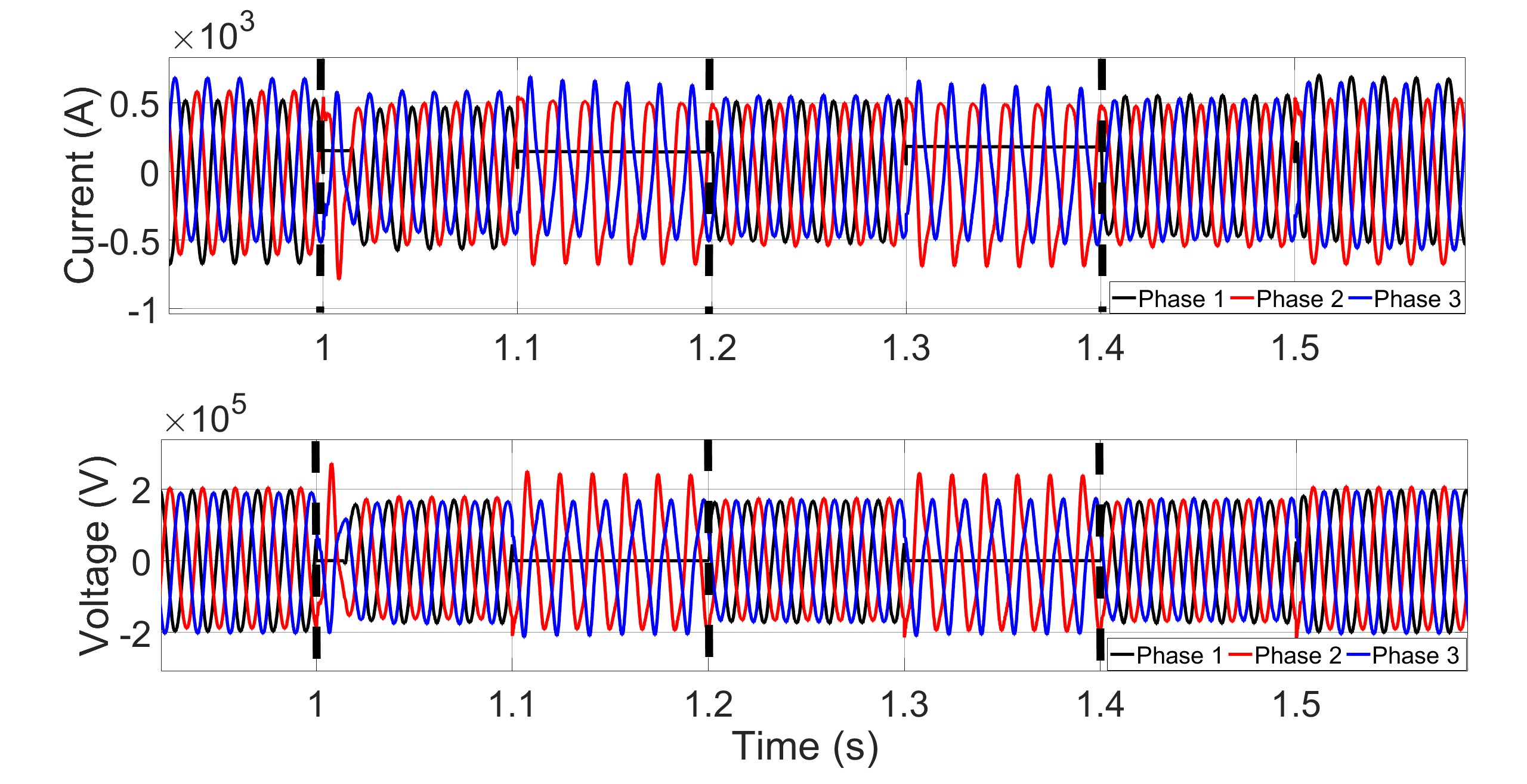}      \vspace{-3mm}
    \caption{Voltage and current waveforms at the \ac{MG} (Bus 24) during switching attacks happening at $0.1$\,s intervals from $t=1$\,s to $t=1.5$\,s.} 
    \label{fig:islanding_waveforms_2}\vspace{-2mm}
\end{figure}

Fig.~\ref{fig:islanding_waveforms_2} shows the three-phase voltage and current waveforms at bus 24 during this attack. Vertical dashed lines denote islanding transitions that begin at $t=1$\,s, with rapid cycling until the final reconnection at $t=1.5$\,s. In islanded mode, voltage and current remain sinusoidal across all phases, though with reduced magnitudes. Upon grid reconnection, Phase 1 voltage and current drop to zero due to a fault, redirecting power flow to ground. Phases 2 and 3 compensate with increased peak currents to account for Phase 1. 
Phase voltage asymmetries could potentially lead to overloading, thereby increasing system losses and reducing reliability. Prolonged operation under such imbalances can cause excessive heating, accelerate equipment degradation, and raise the likelihood of premature failure of grid components and insulation.

\begin{figure}[t]
    \centering
    \includegraphics[width=0.45\textwidth]{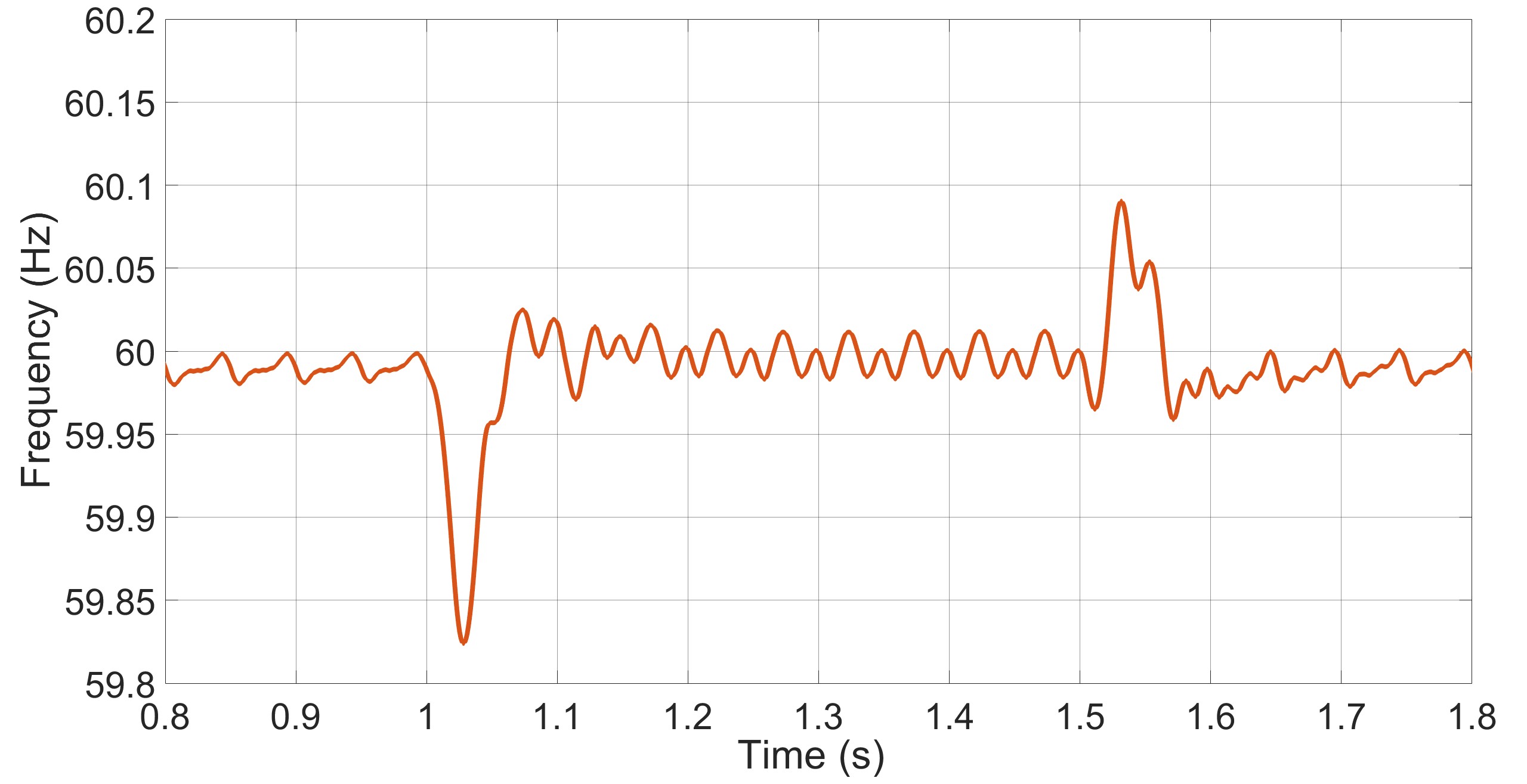}       \vspace{-3mm}
    \caption{Frequency response at the \ac{MG} (Bus 24) during forced islanding at $t=1$\,s and reconnection at $t=1.5$\,s.}
    \label{fig:frequency_response_3}
\end{figure}

\subsubsection{System II -- \ac{IBR}-dominated \ac{MG} Generation}
The following scenarios examine the stability of the \ac{MG} with a 70\% \ac{IBR} and 30\% synchronous generation mix.
\paragraph{Single Forced Islanding Scenario}

In this scenario, the attacker switches the \ac{CB} twice: once to island the system at $t=1$\,s, and once to reconnect it at $t=1.5$\,s. Fig.~\ref{fig:frequency_response_3} shows the frequency response measured at the \ac{MG} at bus 24 during these two key events. Following islanding, the frequency exhibits a more pronounced dip compared to \emph{System I} due to the reduced contribution of synchronous generation. While the frequency remains within acceptable limits, the lower system inertia makes it more vulnerable to sudden disturbances. Between $t=1$\,s and $t=1.5$\,s, the frequency oscillates around 60\,Hz, similar to \emph{System I}, with no significant degradation in stability. After reconnection at $t=1.5$\,s, the initial frequency spike is comparable to that in \emph{System I}, but the system takes slightly longer to stabilize in the grid-connected state. Nonetheless, the frequency settles near 60\,Hz within approximately one second, confirming that the \ac{MG} maintains sufficient control capability even with reduced synchronous support.

\paragraph{\ac{CB} Switching Attack Scenario} 

\begin{figure}[t]
    \centering   \includegraphics[width=0.45\textwidth]{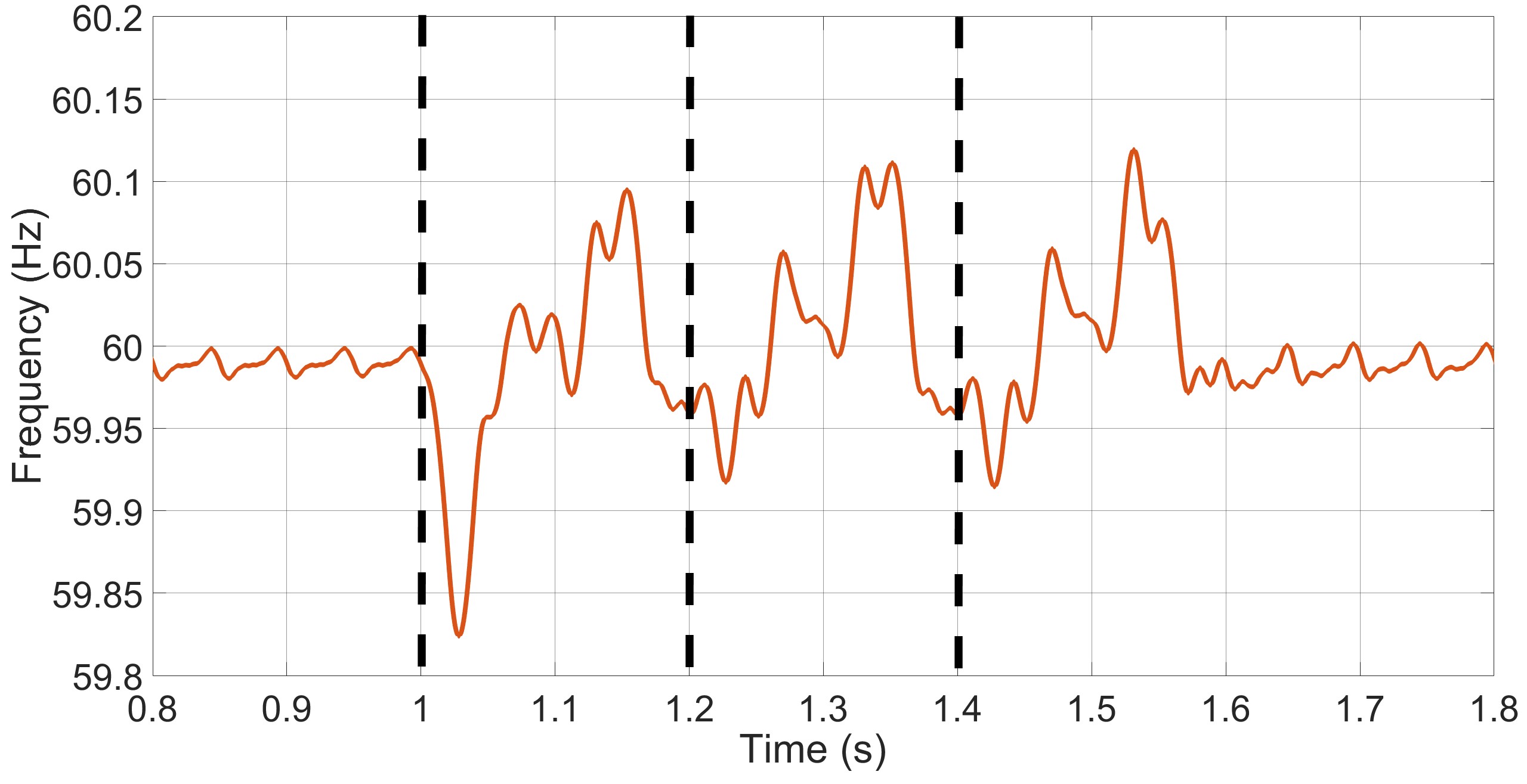}       \vspace{-3mm}
    \caption{Frequency response at the \ac{MG} (Bus 24) during switching attacks happening at $0.1$\,s intervals from $t=1$\,s to $t=1.5$\,s.}
    \label{fig:frequency_response_4} \vspace{-3mm}
\end{figure}

The attacker triggers the \ac{CB} three times within a 0.5-second window. The \ac{CB} is opened every 0.2 seconds between $t=1.0$\,s and $t=1.5$\,s and is re-closed 0.1 seconds later each time. Fig.~\ref{fig:frequency_response_4} shows the resulting frequency response. Each \ac{CB} opening causes a sharp frequency dip followed by a spike upon reconnection. However, the reduced inertia from the lower synchronous generation results in more pronounced frequency deviations. The first islanding event at $t=1$\,s causes the frequency to dip below 59.85\,Hz. Unlike \emph{System I}, the frequency during islanded intervals does not become progressively more stable with each successive event. Instead, the frequency spikes upon reconnection become increasingly pronounced, reaching nearly 60.15\,Hz after the final reconnection. Despite these variations, the frequency remains within the prescribed OF1 and UF1 bounds of \cite{b10} throughout the attack sequence and shows no signs of critical instability before or after these events. \looseness = -1

\subsection{Voltage Stability of \ac{MG} During Islanding}

\begin{figure}[t]
    \centering
    \includegraphics[width=0.45\textwidth]{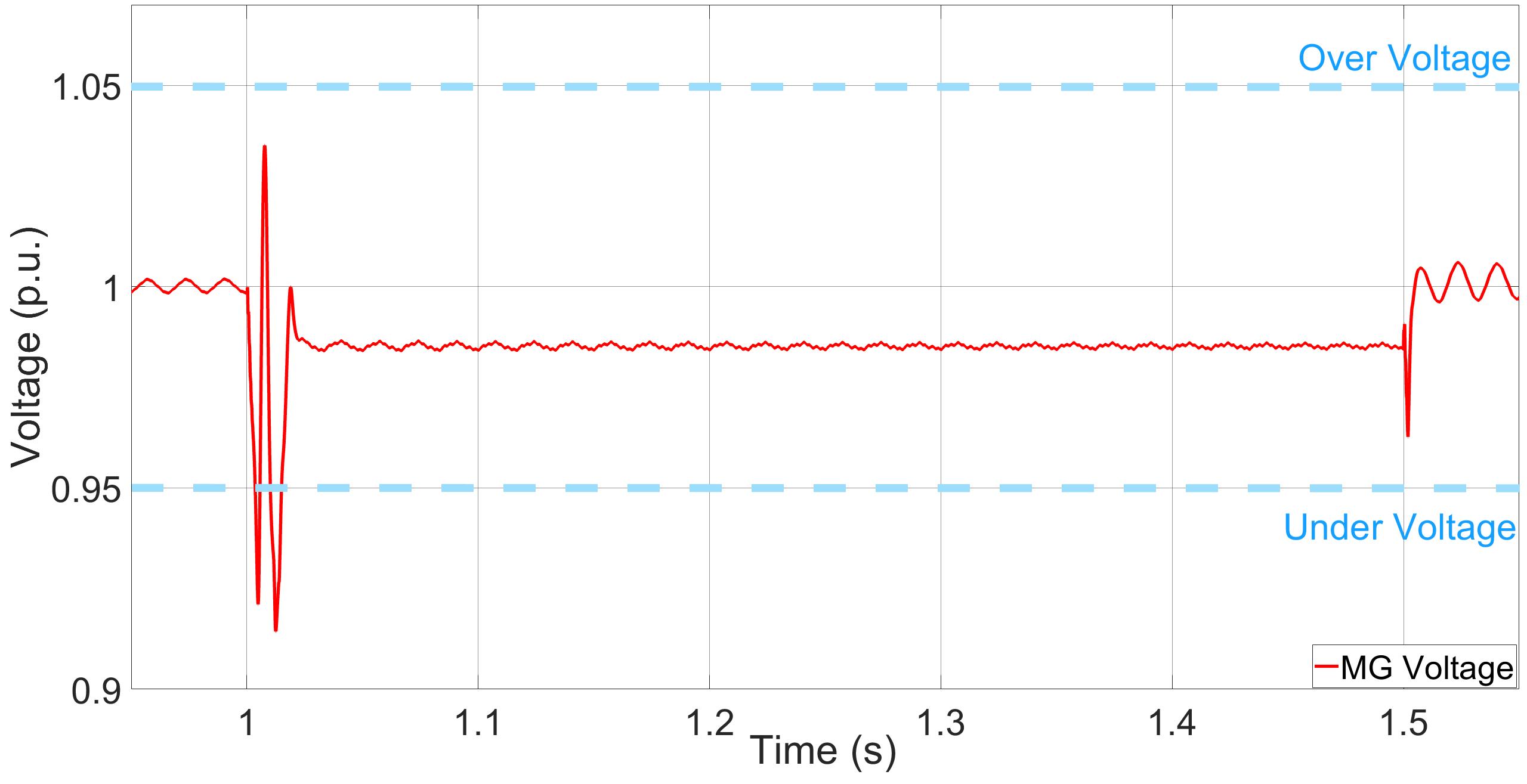}       \vspace{-3mm}
    \caption{Voltage magnitude p.u. at the \ac{MG} (Bus 24) during forced islanding at $t=1$\,s and reconnection at $t=1.5$\,s.} 
    \label{fig:case150onepu}
\end{figure}

\begin{figure}[t]
    \centering
    \includegraphics[width=0.45\textwidth]{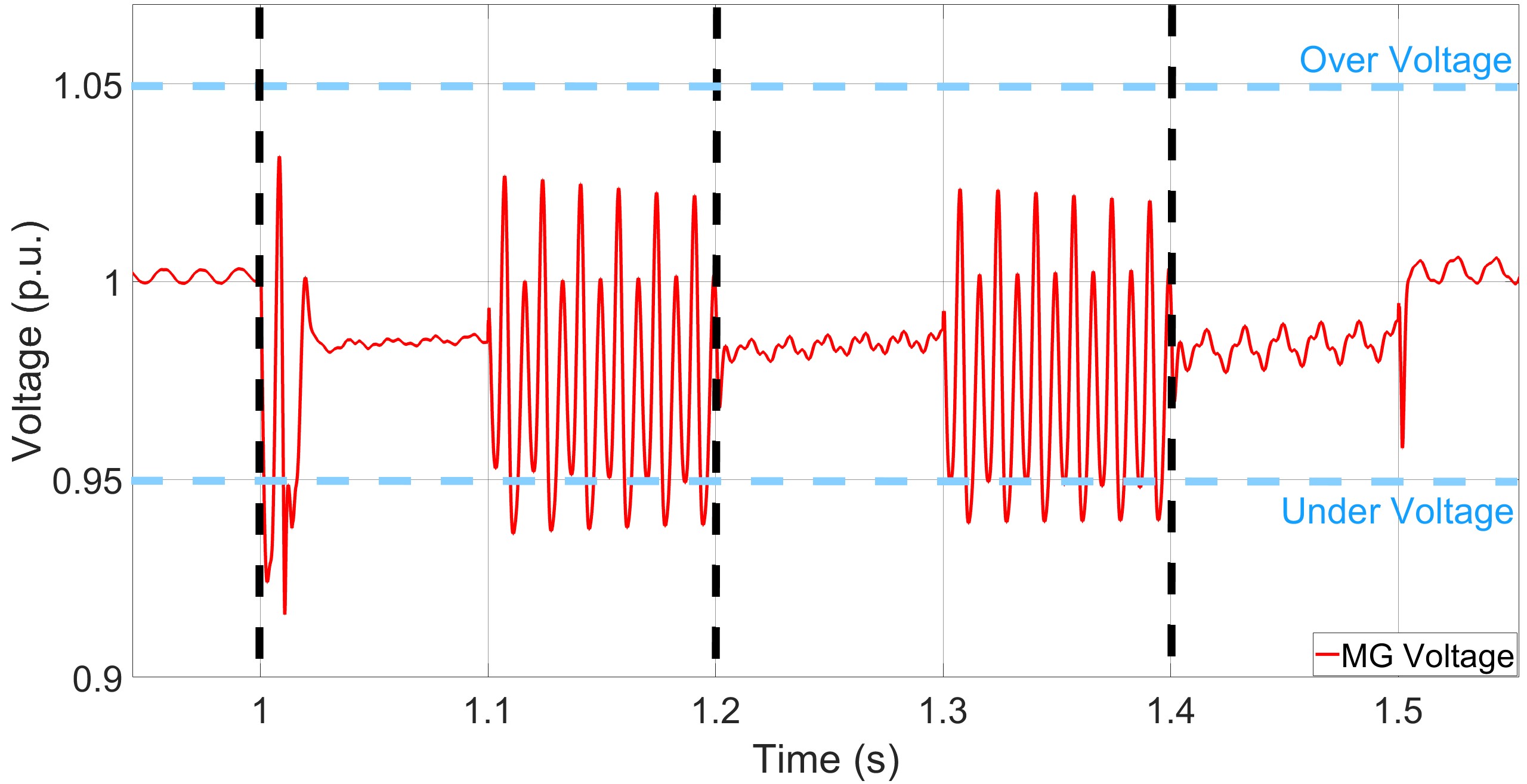}       \vspace{-3mm}
    \caption{Voltage magnitude p.u. at the \ac{MG} (Bus 24) during switching attacks happening at $0.1$\,s intervals from $t=1$\,s to $t=1.5$\,s.} 
    \label{fig:case150rapidpu} \vspace{-3mm}
\end{figure}

\begin{figure}[t]
    \centering
    \includegraphics[width=0.49\textwidth]{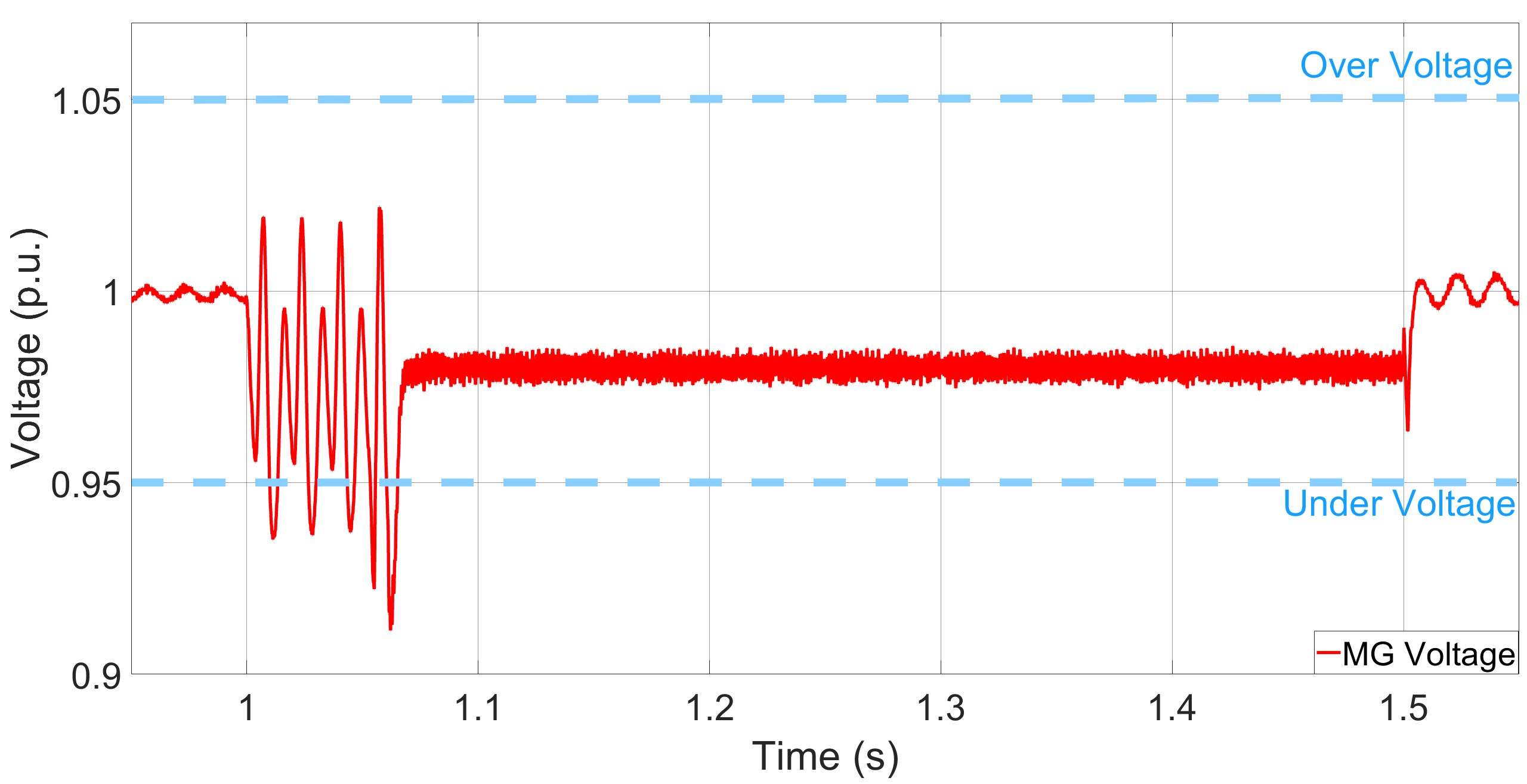}       \vspace{-5mm}
    \caption{Voltage magnitude p.u. at the \ac{MG} (Bus 24) during forced islanding at $t=1$\,s and reconnection at $t=1.5$\,s.} 
    \label{fig:case210onepu}
\end{figure}

\begin{figure}[t]
    \centering
    \includegraphics[width=0.49\textwidth]{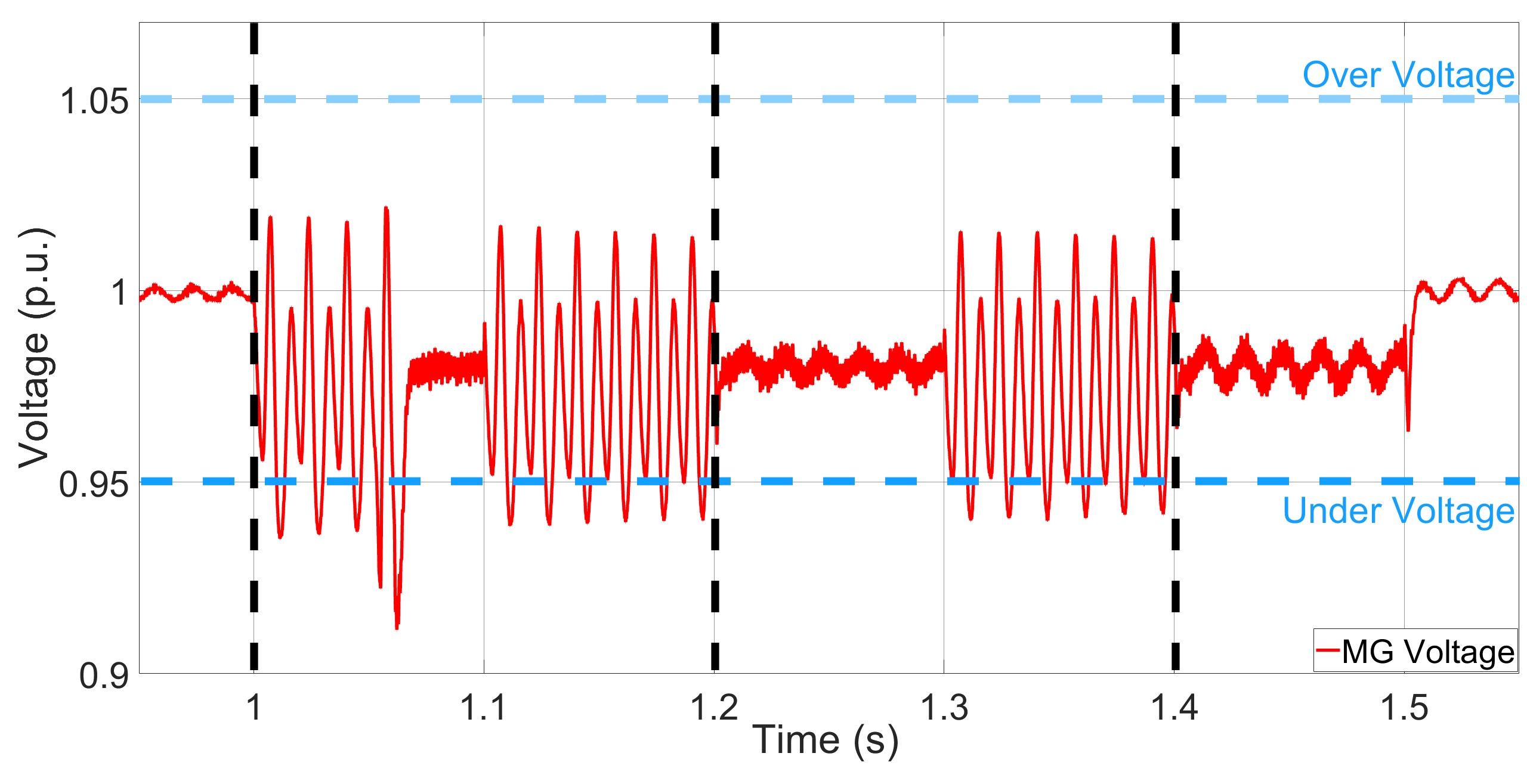}       \vspace{-5mm}
    \caption{Voltage magnitude p.u. at the \ac{MG} (Bus 24) during switching attacks happening at $0.1$\,s intervals from $t=1$\,s to $t=1.5$\,s.} 
    \label{fig:case210rapidpu} \vspace{-3mm}
\end{figure}

Fig.~\ref{fig:case150onepu}, ~\ref{fig:case150rapidpu}, ~\ref{fig:case210onepu}, and ~\ref{fig:case210rapidpu} illustrate the \ac{MG} voltages in both scenarios for each of the two systems described in Table~\ref{tab:scenarioInfo}. As discussed in \cite{MG_pu_voltages}, the typical \ac{MG} voltage limits of 0.95 and 1.05 p.u. are denoted in each Fig. using blue dotted lines.

Fig.~\ref{fig:case150onepu} illustrates the \ac{MG} voltage in \emph{System I, Scenario 1}. At the moment of forced islanding (1 second), the voltage drops below 1 p.u. and then rebounds above 1 p.u. MG voltage exceeds the undervoltage limits of 0.95 p.u. at the moment of islanding, indicating potential transient instability due to the fault existing in the system. For the duration of the islanding, the voltage remains within nominal values, indicating a stable \ac{MG}. Upon reconnection (1.5 seconds), there is a transient dip, followed by stabilization to 0.97 p.u. within 0.01 seconds.  

Fig.~\ref{fig:case150rapidpu} illustrates the \ac{MG} voltage in \emph{System I, Scenario 2}. At the moment of the first forced islanding (1 second), the voltage dips and rebounds similarly to Fig.~\ref{fig:case150onepu}. During forced grid connection, seen after the vertical dashed lines, the \ac{MG} voltage experiences instability, oscillating around 1 p.u. After the first islanded condition, the following islanded conditions show no transient spikes and resemble the nominal grid-connected conditions, showcasing a stable islanded \ac{MG}. However, at the moments of reconnection (i.e., 1.1 seconds) the voltage momentarily drops below the 0.95 p.u. limit, causing instability in the \ac{MG}, as it is forced to reconnect. The final grid connection (1.5 seconds) occurs as the fault clears and the \ac{MG} voltage returns to nominal operation.

Fig.~\ref{fig:case210onepu} illustrates the \ac{MG} voltage in \emph{System II, Scenario 1}. At the moment of forced islanding (1 second), the voltage drops below 1 p.u. and then rebounds above 1 p.u. (similar to Fig.~\ref{fig:case150onepu}). However, the oscillations in Fig.~\ref{fig:case210onepu} are retained for almost double the duration (as opposed to Fig.~\ref{fig:case150onepu}) and exhibit higher voltage drops. Furthermore, the \ac{MG} voltage appears noisier, which is mainly attributed to the inverter's inability to regulate the voltage level. Following the transient spikes, the voltage stabilizes, and upon reconnection at 1.5 seconds, the voltage is brought to 0.97 p.u. within 0.06 seconds. Overall, this indicates a stable \ac{MG} despite the attack; however, in this scenario, the duration and amplitude of the transient spikes are increased with increased \ac{IBR} generation. 

Finally, Fig.~\ref{fig:case210rapidpu} illustrates the \ac{MG} voltage in \emph{System II, Scenario 2}. At the moment of the first forced islanding (1 second), the voltage dips and rebounds similarly to Fig.~\ref{fig:case150rapidpu}, except with prolonged and more severe magnitude transients. 
However, the transient spikes during grid-connected attacking conditions are more severe, as shown by increased peak and trough magnitudes. Additionally, during \ac{MG} islanding, the voltage operates below 1 p.u. for a prolonged duration and exhibits lower values than in Fig.~\ref{fig:case150rapidpu}. Furthermore, the \ac{MG} voltage exceeds the \ac{UV} limits during reconnection. \looseness=-1 


\begin{table}[t!]
\small
\setlength{\tabcolsep}{1.2pt}
\centering
\caption{Summary of System Responses Under Cyberattack Scenarios}
\label{tab:summaryResults}

\renewcommand{\tabularxcolumn}[1]{m{#1}}

\begin{tabularx}{\linewidth}{ 
  || >{\hsize=0.8\hsize\raggedright\arraybackslash}X 
  | >{\hsize=1.2\hsize\centering\arraybackslash}X 
  | >{\hsize=0.75\hsize\centering\arraybackslash}X 
  | >{\hsize=1.25\hsize\raggedright\arraybackslash}X || }

\hline\hline
\textbf{Test Case} & \textbf{Frequency} & \textbf{Voltage} & \textbf{Key Observations} \\
\hline
\textbf{System I, Scenario 1} & Minor dips during switching, fast recovery & Stable & Minimal disruption during islanding and reconnection \\
\hline
\textbf{System I, Scenario 2} & Oscillations with deeper troughs during switching & Slight \ac{UV} & Switching causes compounding instability \\
\hline
\textbf{System II, Scenario 1} & Larger transients & Longer time to stabilize & More \ac{PV} increases frequency/voltage disturbances \\
\hline
\textbf{System II, Scenario 2} & Deep oscillations & Severe \ac{UV} & Most unstable case with high-risk of phase imbalance \\
\hline\hline

\end{tabularx}
\vspace{-5mm}
\end{table}

Table~\ref{tab:summaryResults} summarizes the frequency and voltage responses for each test case. As the table shows, \ac{PV} penetration levels and number of \ac{CB} switches make the system more vulnerable to disturbances. Overall, the results confirm that the \ac{MG} voltage stays within acceptable operational limits across all test scenarios during islanding, demonstrating stable behavior during both single and multiple forced islanding events. However, the \ac{MG} operates below the threshold voltage limits when it is forced to reconnect, causing significant voltage spikes across all voltage phases. At the same time, the transition from a balanced 50\%–50\% generation split to a 70\% \ac{IBR}-based \ac{MG} furnishes more severe transient spikes and longer stabilization times when the \ac{MG} transitions between islanded and grid-connected modes.  
The results highlight that increasing \ac{IBR} penetration inherently heightens the \ac{MG}’s susceptibility to abrupt disturbances due to lack of inertia. Such observations could be exploited by threat actors aiming to maximize their attack impact, leveraging improperly secured grid devices (in our case the \ac{CB}) to mount their attacks.

\section{Conclusion} \label{s:conclusion}
This paper investigates the behavior of an integrated \ac{TnD} system under coordinated switching attacks, leading to \ac{MG} islanding. The developed \ac{TnD} system model couples the IEEE 39-bus transmission network with an MG at the distribution level, which uses a mix of \ac{IBR} and synchronous generation. We present real-time measurements, e.g.,  voltage, current, and frequency, during different attack scenarios demonstrating the effects that different \ac{IBR} penetration levels can induce on \ac{MG} behavior. Future work will incorporate additional \ac{IBR} resources such as \ac{BESS} and increasing \ac{PV} penetrations as well as additional attack scenarios, highlighting the impact on stability of increased \ac{IBR} penetration during adverse events.  


\bibliographystyle{IEEEtran}
\bibliography{biblio}
\end{document}